\documentclass[11pt,a4paper]{article}
\usepackage{jheppub}

\pdfoutput=1
\usepackage{bbm} 
\usepackage{amsthm}
\usepackage{graphicx}
\usepackage{epstopdf}
\usepackage{newlfont}
\usepackage{feynmf}
\usepackage{cleveref}
\usepackage{array}
\usepackage{ytableau}
\usepackage{mathtools}
\usepackage{dsfont}
\usepackage{hyperref}

\newcommand{\ed}{\,.}
\newcommand{\ec}{\,,}
\newcommand{\ecq}{\ec\quad}

\newcommand{\arxivref}[1] {\href{http://xxx.lanl.gov/abs/#1}{{\tt #1}}}

\newcommand{\id}[5]{\left(\mathds{1}_{#5}\right)^{#1  ~#3}_{#2 ~#4}}

\newcommand{\cD}{\ensuremath{\mathcal{D}}}

\newcommand{\cG}{\ensuremath{\mathcal{G}}}

\newcommand{\cJ}{\ensuremath{\mathcal{J}}}

\newcommand{\cN}{\ensuremath{\mathcal{N}}}
\newcommand{\cO}{\ensuremath{\mathcal{O}}}

\renewcommand{\l}{\ensuremath{\ell}}

\title{
Bounds on $\cN=1$ Superconformal Theories with Global Symmetries}

\author[a]{Micha Berkooz}
\author[b]{Ran Yacoby}
\author[a]{Amir Zait}

\emailAdd{micha.berkooz@weizmann.ac.il}
\emailAdd{ryacoby@princeton.edu}
\emailAdd{amir.zait@weizmann.ac.il}

\affiliation[a]{
Department of Particle Physics and Astrophysics,
The Weizmann Institute of Science,
Rehovot 76100, Israel
}

\affiliation[b]{
Department of Physics, Princeton University,
Princeton NJ 08544, USA
}

\abstract{
Recently, the conformal-bootstrap has been successfully used to obtain generic bounds on the spectrum and OPE coefficients of unitary conformal field theories. In practice, these bounds are obtained by assuming the existence of a scalar operator in the theory and analyzing the crossing-symmetry constraints of its 4-point function. In $\cN=1$ superconformal theories with a global symmetry there is always a scalar primary operator, which is the top of the current multiplet. In this paper we analyze the crossing-symmetry constraints of the 4-point function of this operator for $\cN=1$ theories with $SU(N)$ global symmetry. We analyze the current-current OPE and write the superconformal blocks, generalizing the work of Fortin, Intriligator and Stergiou to the non-Abelian case. Moreover we find new contributions to the OPE which can appear both in the Abelian and non-Abelian cases. We then use these results to obtain lower bounds on the coefficient of the current 2-point function.
}


\begin{document}

\begin{flushright}
WIS/02/14-FEB-DPPA
\end{flushright}

\maketitle


\setlength{\unitlength}{1mm}

\section{Introduction}

Recently, there has been much interest in generating numerical constraints on conformal field theories using the conformal bootstrap \cite{Rattazzi:2008pe,Rychkov:2009ij,Caracciolo:2009bx, Poland:2010wg,Rattazzi:2010gj,Rattazzi:2010yc, Poland2011,Vichi:2011ux,ElShowk:2012hu,ElShowk:2012ht,Liendo:2012hy,Beem:2013qxa,Kos:2013tga,El-Showk:2013nia,Alday:2013opa,Gaiotto:2013nva}. In these works the existence of a scalar primary operator $\phi$ of dimension $\Delta$ is assumed. Using the conformal bootstrap on the 4-point function of $\phi$, it is then possible to generate numerically bounds on operator dimensions and OPE coefficients of operators in the $\phi\times\phi$ OPE as a function of $\Delta$. The crucial ingredient which allows us to generate these bounds is the knowledge of all the scalar conformal blocks \cite{Dolan:2000ut,Dolan:2003hv}, which encode the dependence of the $4$-point function of $\phi$ on each conformal family in the $\phi\times\phi$ OPE\footnote{More precisely, for doing numerics it is sufficient to have a systematic way to approximate the conformal blocks. For scalar 4-point functions this can be done efficiently in any dimension \cite{ElShowk:2012ht, Kos:2013tga, Hogervorst:2013kva}.}.

It would interesting to apply these methods without introducing any assumption on the operator spectrum. For instance, one would like to analyze the four-point function of the stress-tensor, which exists for any CFT. More generally, assuming the CFT has some global symmetry one would like to understand the constraints of conformal invariance arising from application of the conformal bootstrap to the $4$-point function of the global symmetry current. Unfortunately, to this date there are no closed form expressions for the conformal blocks of non-scalar operators (see however \cite{Costa:2011dw,SimmonsDuffin:2012uy}), so these interesting directions cannot be pursued in a straightforward way yet. 

However, in supersymmetric theories the situation is better since in some cases, symmetry currents reside in multiplets whose superconformal-primary (sprimary) is a scalar field. For instance in $\cN=4$ supersymmetric Yang-Mills the energy-momentum tensor resides in a multiplet whose sprimary is a scalar in the $\mathbf{20'}$ representation of the $SU(4)_R$ $R$-symmetry group. The bootstrap constraints for this case were recently analyzed in \cite{Beem:2013qxa}.

Similarly, in any four-dimensional $\cN=1$ superconformal theory, a global symmetry current $j^a_{\mu}$ resides in a real multiplet 
\begin{align}
\cJ^a(z) = \cJ^a(x,\theta,\bar{\theta}) = J^a(x) + i\theta j^a(x) - i\bar{\theta}\bar{j}^a(x) - \theta\sigma^{\mu}\bar{\theta} j^a_{\mu}(x) + \cdots\ec\label{susyJ}
\end{align}
which satisfies $D^2\cJ^a=\bar{D}^2\cJ^a=0$, and the omitted terms in the equation above are determined by this constraint. The sprimary $J^a$ is a dimension two real scalar field in the adjoint representation of the symmetry group.

In this paper we will use the conformal bootstrap to constrain four dimensional $\cN=1$ superconformal theories with an $SU(N)$ global symmetry\footnote{The generalization to other symmetry groups is straightforward.}. In particular, we will place lower bounds on the current ``central charge'' $\tau$  defined as the coefficient of the current 2-point function\footnote{Similar bounds were generated in \cite{Poland:2010wg, Poland2011}, assuming the existence of a charged scalar primary. Here we assume supersymmetry instead.}
\begin{align}
\langle j_{\mu}^a(x) j_{\nu}^b(0) \rangle =  3 \tau \, \delta^{ab} \, \frac{x^2\delta_{\mu\nu}-2x_{\mu}x_{\nu}}{4\pi^4x^8} \ed \label{jj}
\end{align}

The decomposition of the 4-point function of $J^a$ into conformal blocks is constrained by supersymmetry. In particular, the OPE coefficients in $J^a\times J^b$ of different primary operators in a super-multiplet are not independent and the corresponding conformal blocks are re-packaged into the so-called superconformal blocks. These constraints were already analyzed in detail in \cite{Fortin:2011nq} for the $U(1)$ case, and will generalize those results to the non-Abelian case. In addition, we find new operators which generally appear in the OPE which were not found in \cite{Fortin:2011nq}.

The form of the bounds we find is $\tau > f(N)$. Qualitatively, the existence of a lower bound means there is a minimal amount of ``charged stuff'' which must exist in any such theory. A free chiral superfield has, in our normalization, $\tau=1$. We do not know of any theory with $\tau<1$ and it would be very interesting to understand whether those exist, or alternatively to prove that $\tau\ge 1$ in general. 

The paper is organized as follows.
In section \ref{prelim} we briefly review the conformal bootstrap and set up our conventions. In addition, we determine the sum-rules which result from applying crossing-symmetry to the 4-point function of a scalar primary in the adjoint representation of $SU(N)$. In section \ref{SUSY} we discuss the constraints imposed by $\cN=1$ superconformal invariance on the $J^a\times J^b$ OPE and superconformal blocks. In section \ref{results} we present the lower bounds we obtained on $\tau$ and a short discussion.

\section{Preliminaries}
\label{prelim}
\subsection{Conformal Bootstrap}
In this section we spell out our normalization conventions and briefly summarize the conformal bootstrap constraint for a general CFT. The reader is referred to \cite{Rattazzi:2008pe} for a more extensive treatment.

Consider a general CFT in four Euclidean dimensions, and in particular the subset of operators consisting of spin-$\l$ primary operators $\cO_I^{(\l)} \equiv \cO_I^{\mu_1\cdots\mu_\l}$, which are symmetric-traceless rank-$\l$ tensors (i.e. in the $(\l/2,\l/2)$ representation of the Lorentz group $SO(4)$). The index $I$ labels the primary operators in the CFT, and we will denote the complex conjugate operator by a barred index $\bar{\cO}_{\bar{J}}\equiv(\cO_J)^{\dag}$.

We set the normalization of such operators by demanding that their $2$-point function is of the form 
\begin{align}
\langle \cO_I^{\mu_1\cdots\mu_\l}(x_1) \bar{\cO}_{\bar{J}}^{\nu_1\cdots\nu_\l}(x_2)\rangle  &= \delta_{I\bar{J}} \frac{I^{\mu_1\nu_1}(x_{12})\cdots I^{\mu_\l\nu_\l}(x_{12})}{x_{12}^{2\Delta_I}} \ec \label{2pnt} \\
I^{\mu\nu}(x) &\equiv \delta^{\mu\nu} - 2\frac{x^{\mu}x^{\nu}}{x^2} \ec
\end{align}
where on the RHS the indices $(\mu_1,\ldots,\mu_\l)$ and $(\nu_1,\ldots,\nu_\l)$ should be symmetrized with the traces removed, and $\Delta_I$ denotes the dimension of $\cO_I^{(\l)}$.
The $3$-point function of a spin-$\l$ primary with two scalar primaries $\phi_a$, $\phi_b$ of equal dimension $\Delta_0$ is
\begin{align}
\langle \phi_a(x_1) \phi_b(x_2) \bar{\cO}_{\bar{I}}^{\mu_1\cdots\mu_\l} \rangle &= \lambda_{ab,I} \frac{Z^{\mu_1}\cdots Z^{\mu_\l}}{x_{12}^{2\Delta_0 - \Delta_I + \l} x_{23}^{\Delta_I-\l}x_{31}^{\Delta_I-\l}} \ec \label{3pnt} \\
Z^{\mu} &\equiv \frac{x_{31}^{\mu}}{x_{31}^2} - \frac{x_{32,}^{\mu}}{x_{32}^2} \ec
\end{align}
where again the Lorentz indices on the RHS should be symmetrized with the traces removed, and $a,b$ are arbitrary labels.

The information on the $2$-point and $3$-point functions \eqref{2pnt}, \eqref{3pnt} is contained in the $\phi_a\times\phi_b$ OPE (suppressing Lorentz indices for simplicity),
\begin{align}
\phi_a(x)\phi_b(0) = \frac{\delta_{ab}}{x^{2\Delta_0}}\mathds{1} \,\,+\!\! \sum_{\cO_I=\text{primary}} \lambda_{ab,I} C_{\cO}(x,\partial)\cO_I \ed
\end{align}
In above equation the identity operator $\mathds{1}$
contains the information on the $2$-point function and the sum over primaries encodes all the information on $3$-point functions. The operator $C_{\cO}(x,\partial)$ is entirely determined by conformal symmetry, and encodes the contributions of all the descendants of $\cO_I$ to the OPE.

In a unitary theory if the scalars $\phi_a$ are real but $\cO_I^{(\l)}$ are complex then the OPE coefficients are generally complex and satisfy $\lambda_{ab,\bar{I}} = (\lambda_{ab,I})^*$. If we choose a real basis of operators, then the OPE coefficients $\lambda_{ab,I}$ must be real $\lambda_{ab,\bar{I}} = \lambda_{ab,I}$. The $3$-point function is non-zero only for $\cO_I$ of integer spin, and (odd) even spins correspond to the (anti-)symmetric combination of $\phi_a$ and $\phi_b$ (i.e. $\lambda_{ab,I}$ is \mbox{(anti-)symmetric} in $a,b$ for (odd) even spins).

The crossing-symmetry constraints for the 4-point function $\langle \phi_{a}(x_1)\phi_{b}(x_2)\phi_{c}(x_3)\phi_{d}(x_4)\rangle$ are obtained by using the OPE in the $(12)(34)$ (``$s$-channel'') and $(14)(23)$ (``$t$-channel'') channels and equating the results,
\begin{align}
u^{-\Delta_0} \cdot \!\!\!\! \sum_{\cO_I \in \phi_{a}\times\phi_{b}} \frac{\lambda_{a b, I}\lambda_{c d, \bar{I}}}{(-2)^{\l_I}} g_{\Delta_I,\l_I}(u,v) = v^{-\Delta_0} \cdot \!\!\!\! \sum_{\cO_I \in \phi_{a}\times\phi_{d}} \frac{\lambda_{a d, I}\lambda_{b c, \bar{I}}}{(-2)^{\l_I}} g_{\Delta_I,\l_I}(v,u) \ec \label{sumrule}
\end{align}
whered $g_{\Delta,\l}$ are the scalar conformal blocks \cite{Dolan:2000ut,Dolan:2003hv},
\begin{align}
g_{\Delta, \l}(u, v)&=\frac{z\bar z}{z-\bar z}\left( k_{\Delta +\ell}(z)k_{\Delta -\l -2}(\bar z)-(z\leftrightarrow \bar z)\right) \label{cb} \ec\\
k_{\beta} (x)&\equiv x^{\beta /2}{}_2F_{1}(\beta /2, \beta/2, \beta ; x) \ec
\end{align}
and
\begin{align}
u = \frac{x_{12}^2 x_{34}^2}{x_{13}^2 x_{24}^2} = z\bar{z} \ecq
v = \frac{x_{14}^2 x_{23}^2}{x_{13}^2 x_{24}^2} = (1-z)(1-\bar{z}) \label{uvCR}\ec
\end{align}
are the two conformal cross-ratios. In \eqref{sumrule} the summation is over all primary operators in the $\phi\times\phi$ OPE\footnote{If an operator is complex then its complex conjugate should also be included in the sum as an independent primary operator.
}.

\subsection{Bootstrap for Scalars in the Adjoint of $SU(N)$}
\label{adjScalar}
In this section we discuss a specific case of the general bootstrap constraint \eqref{sumrule} in which $\phi_a$ is a real scalar primary in the adjoint representation\footnote{The indices $a,b,\ldots,$ label the adjoint representation of $SU(N)$ and $i,j,\ldots,$ are (anti-)fundamental indices. We will sometimes find it more convenient to work in the fundamental basis with $\phi^m_{~i}\equiv \phi^a(T^a)^m_{~i}$, where $(T^a)^m_{~i}$ is a generator in the fundamental of $SU(N)$. Our normalization convention is $(T^a)^m_{~i}(T^a)^n_{~j} = \delta_i^n\delta_j^m - \frac{1}{N}\delta_i^m\delta_j^n$. The structure constants are $i f^{abc} = \text{tr}\left([T^a,T^b]T^c\right)$ and $d^{abc} = \text{tr}\left(\{T^a,T^b\}T^c\right)$.} of $SU(N)$. We will later apply the results  of this section to the case in which this scalar is the top of the current multiplet in $\cN=1$ theories. The crossing-symmetry relations in CFTs with global symmetries were considered in full generality in \cite{Rattazzi:2010yc}, and we apply these results to our case of interest.

The operators which appear in the $\phi_a\times\phi_b$ OPE can be decomposed into any of the $7$ irreducible representations in the product of two adjoint representations of $SU(N)$. Each such representation arises from either the symmetric or anti-symmetric product. Operators in the $\phi_a\times\phi_b$ OPE which are in a (anti-)symmetric representation must be of (odd) even spin from Bose symmetry. The reader is referred to appendix \ref{group} for details regarding the tensor product of two $SU(N)$ adjoint representations and our notations.

Let $\cO^{\mathbf{r}}_I$ be an operator in representation $\mathbf{r}$ which appears in the $\phi\times\phi$ OPE, with $I,J,\ldots,$ labeling the elements of the representation. We denote the corresponding OPE coefficient (defined in \eqref{3pnt}) as $\lambda_{a b\ec I}^{\cO_{\mathbf{r}}}$ and split it to a universal group factor times some coefficient,
\begin{align}
\lambda_{a b, I}^{\cO_{\mathbf{r}}} &\equiv \lambda_{\cO} C_{ab,I}^{(\mathbf{r})} \ec \label{OPEcf}\\
\lambda_{a b, \bar{I}}^{\bar{\cO}_{\bar{\mathbf{r}}}} &= \lambda_{\bar{\cO}} \bar{C}_{ab,\bar{I}}^{(\bar{\mathbf{r}})} \equiv (\lambda_{\cO} C_{ab,I}^{(\mathbf{r})})^* \ec
\end{align}
where $C_{ab,I}^{(\mathbf{r})}$ is the relevant Clebsch-Gordan coefficient, and is the same for any operator in the representation $\mathbf{r}$, while the coefficient $\lambda_{\cO}$ is the same for each element of the representation.
The sum-rule in \eqref{sumrule} becomes
\begin{align}
u^{-\Delta_0} &\sum_{\mathbf{r}} \sum_{\cO \in (\phi\times\phi)_{\mathbf{r}}} \frac{|\lambda_{\cO}|^2}{(-2)^{\l_{\cO}}} \delta^{I\bar{J}} C_{ab,I}^{(\mathbf{r})}\bar{C}_{cd,\bar{J}}^{(\bar{\mathbf{r}})} \, g_{\Delta_{\cO},\l_{\cO}}(u,v) \nonumber \\
&= v^{-\Delta_0} \sum_{\mathbf{r}} \sum_{\cO \in (\phi\times\phi)_{\mathbf{r}}} \frac{|\lambda_{\cO}|^2}{(-2)^{\l_{\cO}}} \delta^{I\bar{J}} C_{ad,I}^{(\mathbf{r})}\bar{C}_{bc,\bar{J}}^{(\bar{\mathbf{r}})} \, g_{\Delta_{\cO},\l_{\cO}}(v,u) \ed \label{SR1}
\end{align}

Each term in the second sums in equation \eqref{SR1} has the same sign from the $(-2)^\l$ factor, as this only depends on whether $\mathbf{r}$ is in the symmetric or anti-symetric product of two adjoints\footnote{The adjoint representation appears both in the symmetric and anti-symmetric product and we count those as distinct in the sum over representations in \eqref{SR1}.}. We use this property to write the above sum-rule as
\begin{gather}
u^{-\Delta_0} \sum_{\mathbf{r}} (\mathds{1}_{\mathbf{r}})_{ab,cd} G_{\mathbf{r}}(u,v)  = v^{-\Delta_0} \sum_{\mathbf{r}} (\mathds{1}_{\mathbf{r}})_{ad,cb} G_{\mathbf{r}}(v,u) \ec \notag\\
G_{\mathbf{r}}(u,v) \equiv \! \sum_{\cO\in(\phi\times\phi)_{\mathbf{r}}} p_{\cO}\, g_{\Delta_{\cO},\l_{\cO}}(u,v) \ecq 
p_{\cO} \equiv \frac{|\lambda_{\cO}|^2}{2^{\l_{\cO}}}\ed \label{SUnSR}
\end{gather}

where $G_{\mathbf{r}}(u,v)$ is the sum over conformal blocks in a given representation, and $(\mathds{1}_{\mathbf{r}})_{ab,cd}=\pm \delta^{I\bar{J}} C_{ab,I}^{(\mathbf{r})}\bar{C}_{cd,\bar{J}}^{(\bar{\mathbf{r}})}$ is just the identity matrix in the representation $\mathbf{r}$ projected to adjoint representation indices up to a sign, which can be determined by reflection positivity. Explicit expressions for these identity matrices are given in \eqref{id}.

After plugging \eqref{id} into the sum-rule \eqref{SUnSR}, it can be decomposed into several equations by equating the coefficients of the independent delta-functions in the identity matrices. We do this in the next subsections paying attention to the special cases $SU(2)$ and $SU(3)$. 

The resulting sum-rules are conveniently expressed in terms of the functions
\begin{align}
F_{(\mathbf{r})}(u,v) &\equiv \sum_{\cO\in(\phi\times\phi)_{(\mathbf{r})}} \!\!\! p_{\cO} \,F_{\Delta_{\cO},\l_{\cO}}(u,v) \equiv \frac{u^{-\Delta_0}G_{(\mathbf{r})}(u,v) - v^{-\Delta_0}G_{(\mathbf{r})}(v,u)}{v^{-\Delta_0}-u^{-\Delta_0}} \ec \label{eq:F} \\
H_{(\mathbf{r})}(u,v) &\equiv \sum_{\cO\in(\phi\times\phi)_{(\mathbf{r})}} \!\!\! p_{\cO} \,H_{\Delta_{\cO},\l_{\cO}}(u,v) \equiv \frac{u^{-\Delta_0}G_{(\mathbf{r})}(u,v) + v^{-\Delta_0}G_{(\mathbf{r})}(v,u)}{v^{-\Delta_0}+u^{-\Delta_0}} \label{eq:G} \ed
\end{align}

We have verified that all the sum-rules written below are obeyed by both the four point function of a free scalar field in the adjoint of $SU(N)$, and that of the adjoint bilinear $\bar{\phi}^i (T^a)_i^j\phi_j$ in the theory of a free fundamental scalar $\phi_i$. 

\subsubsection{$U(1)$}
\label{U1}
For $U(1)$ the sum-rule is the usual one for the 4-point function of a real scalar operator:
\begin{align}
\label{sum_rule_u1}
F(u,v)=\sum_{\cO\in(\phi\times\phi)_{(1)}} \!\!\! p_{\cO} \,F_{\Delta_{\cO},l_{\cO}}(u,v)=1\ec
\end{align}
where we separated the contribution of the identity operator for which $p_{\mathds{1}}=1$ and  $g_{0,0}=F_{0,0} = -1$. 

\subsubsection{$SU(2)$}
\label{SU2}
For $SU(2)$ we have $\mathbf{3}\times \mathbf{3} = \mathbf{5}_s + \mathbf{3}_a + \mathbf{1}_s$ corresponding to the representations $(S,\bar{S})_s$, $(Adj)_a$ and the trivial representation. Setting all the terms which correspond to the other representations in \eqref{SUnSR} to zero, plugging in the expressions for the identity matrices \eqref{id} and equating independent coefficients, we can express the result as three independent sum-rules\footnote{Equivalent sum-rules were also worked out in \cite{Rattazzi:2010yc} for scalars in the fundamental of $SO(3)$. Our result is consistent with \cite{Rattazzi:2010yc}, but we work in slightly different convention such that $G_1^{here}=2G_1^{there}$, which amounts to a rescaling of all the OPE coefficients in the trivial representation by a factor of $2$.},
\begin{align}
F_{(S,\bar{S})_s} - F_{Adj_a} &= 0 \ec \label{eq:SU2sr1}\\
\frac{2}{3}F_{(S,\bar{S})_s} + 2F_{Adj_a} + F_1 &= 1 \label{eq:SU2sr2}\ec \\
\frac{10}{3}H_{(S,\bar{S})_s} + 2H_{Adj_a} - H_1 &= 1  \label{eq:SU2sr3}\ed
\end{align}

\subsubsection{$SU(3)$}
\label{SU3}
For $SU(3)$ the $(A,\bar{A})_s$ representation does not exist so we set it to zero in \eqref{SUnSR}. The resulting sum-rules are given by
\begin{gather}
\frac{3}{2}F_{(S,\bar{S})_s} + F_{(S,\bar{A})_a} + F_1 = 1 \ec \\
\frac{9}{5}F_{(S,\bar{S})_s} - \frac{3}{2}F_{(S,\bar{A})_a} + F_{(Adj)_s} = 0 \ec \\
F_{(S,\bar{S})_s} - \frac{1}{6}F_{(S,\bar{A})_a} - F_{(Adj)_a} = 0 \ec \\
\frac{9}{10} H_{(S,\bar{S})_s} + H_{(S,\bar{A})_a} + \frac{4}{3}H_{(Adj)_s} - H_1 = 1 \ec \\
\frac{5}{2}H_{(S,\bar{S})_s} + \frac{5}{9}H_{(S,\bar{A})_a} + \frac{4}{3}H_{(Adj)_a} - H_1 = 1\ed
\end{gather}

\subsubsection{$SU(N)$ for $N>3$}
\label{SUN}
For $N>3$ all the 7 representations listed in appendix \ref{group} can appear in the OPE, and we find
\begin{gather}
F_{(S,\bar{S})_s} + F_{(A,\bar{A})_s} - F_{(S,\bar{A})_a} = 0 \ec \\
\frac{1}{N+2} F_{(S,\bar{S})_s} - \frac{1}{N-2} F_{(A,\bar{A})_s} + \frac{2}{N} F_{(Adj.)_s} = 0 \ec \\
\frac{1}{N+2} F_{(S,\bar{S})_s} + \frac{1}{N-2} F_{(A,\bar{A})_s} + \frac{1}{N} F_{(S,\bar{A})_a} - F_{(Adj.)_s} - F_{(Adj.)_a} = 0 \ec \\
\frac{2N^2}{N^2+3N+2} F_{(S,\bar{S})_s} + \frac{2N^2}{N^2-3N+2} F_{(A,\bar{A})_s} - \frac{16}{N} F_{(Adj.)_s} + F_{(\mathds{1})_s} = 1 \ec \\
\frac{N(N+3)}{N^2+3N+2} H_{(S,\bar{S})_s} + \frac{N(N-3)}{N^2-3N+2} H_{(A,\bar{A})_s} + H_{(S,\bar{A})_a} + \frac{4}{N} H_{(Adj.)_s} - H_{(\mathds{1})_s} = 1 \ec \\
\frac{N+3}{N+2} H_{(S,\bar{S})_s} - \frac{N-3}{N-2} H_{(A,\bar{A})_s} - \frac{1}{N} H_{(S,\bar{A})_a} - H_{(Adj.)_s} + H_{(Adj.)_a} = 0 \ed
\end{gather}

\section{Conformal Bootstrap for Conserved Currents in $\cN=1$ SCFTs}
\label{SUSY}

Consider an $\cN=1$ superconformal field theory with  global symmetry group $\cG$. 
In this section we will analyze the bootstrap constraints for the $4$-point function of $J^a(x)$, which is the top of the current multiplet $\cJ^a(z)$ defined in \eqref{susyJ}. In particular, we extend the results of \cite{Fortin:2011nq} for $U(1)$ to the non-Abelian case, and also find additional possible operators in the $J^a\times J^b$ OPE. We use the notations and conventions of \cite{Fortin:2011nq}. 

\subsection{Current-Current OPE in $\cN=1$ SCFTs}

The general form of the 3-point function of sprimary operators was found in \cite{Osborn:1998qu}. For the 3-point function of two conserved currents with some other sprimary $\cO$ in some representation $\mathbf{r}$ the result is
\begin{align}
\langle \cJ_a(z_1) \cJ_b(z_2) \cO^i_I(z_3) \rangle =
C_{ab,I}^{(\mathbf{r})} \frac{t^i(X,\Theta,\bar{\Theta})}{x_{\bar{1}3}^2 x_{\bar{3}1}^2 x_{\bar{2}3}^2 x_{\bar{3}2}^2}\ec	
\label{JJO1}
\end{align}
where the superspace coordinates are $z_j=(x_j,\theta_j,\bar{\theta}_j)$, and we define 
\begin{equation}
x_{\bar{i}j}^{\mu} = -x_{j\bar{i}} \equiv x_{ij}^{\mu} - i\theta_i\sigma^{\mu}\bar{\theta}_j + i\theta_j\sigma^{\mu}\bar{\theta}_i - i\theta_{ij}\sigma^{\mu}\bar{\theta}_{ij}.
\end{equation}
The quantities $X$, $\Theta$ and $\bar{\Theta}$ are functions of the superspace coordinates given by 
\begin{align}
X &\equiv \frac{x_{3\bar{1}} \tilde{x}_{\bar{1}2} x_{2\bar{3}}}{x_{\bar{1}3}^2 x_{\bar{3}2}^2}\ec &
\Theta &\equiv i\left(\frac{1}{x_{\bar{1}3}^2} x_{3\bar{1}} \bar{\theta}_{31} - \frac{1}{x_{\bar{2}3}^2} x_{3\bar{2}} \bar{\theta}_{32}\right)\ec &
\bar{\Theta} &= \Theta^{\dag}, 
\end{align}
and $i$ labels the Lorentz representation of $\cO$. 

The function $t^i(X,\Theta,\bar{\Theta})$ has to scale appropriately with respect to dilatations and $U(1)_R$ transformations\footnote{The scaling is $t(\lambda\bar{\lambda}X,\lambda\Theta,\bar{\lambda}\bar{\Theta})=\lambda^{2a}\bar{\lambda}^{2\bar{a}}t(X,\Theta,\bar{\Theta})$, where $a-2\bar{a}=2-q$, and $\bar{a}-2a=2-\bar{q}$. The R-charge and dimension of $\cO$ are related to $(q,\bar{q})$ by $
R_{\cO} = \frac{2}{3}(q-\bar{q})$ and $\Delta_{\cO} = q+\bar{q}
$.}. Moreover, because of current conservation, $D^2\cJ^a=\bar{D}^2\cJ^a=0$, the correlator \eqref{JJO1} satisfies a differential equation. As shown in \cite{Osborn:1998qu}, this equation can be translated to the following differential equation for $t$:
\begin{align}
\cD^2 t = \bar{\cD}^2 t = 0 \ec \label{Dsquare}
\end{align}
where
\begin{align}
\cD_{\alpha} = \frac{\partial}{\partial\Theta^{\alpha}}-2i\left(\sigma^{\rho}\bar{\Theta}\right)
_{\alpha}\frac{\partial}{\partial X^{\rho}} \ecq
\bar{\cD}_{\dot{\alpha}} = -\frac{\partial}{\partial\bar{\Theta}^{\dot{\alpha}}} \ed 
\end{align}

Note that $\eqref{JJO1}$ is symmetric under $(z_1,a)\leftrightarrow (z_2,b)$. Therefore, since $C_{ab,I}^{(\mathbf{r})}$ is either symmetric or anti-symmetric under $a\leftrightarrow b$, we need to find $t^i(X,\Theta,\bar{\Theta})$ which is either symmetric or anti-symmetric under $z_1\leftrightarrow z_2$. Under $z_1 \leftrightarrow z_2$ we have $(X,\Theta,\bar{\Theta})\leftrightarrow (-\bar{X},-\Theta,-\bar{\Theta})$, with $\bar{X}^{\mu} = X^{\mu} + 2 i \Theta\sigma^{\mu}\bar{\Theta}$.
It is therefore useful to define
\begin{align}
X_{+}^{\mu} &= \frac{1}{2}(X^{\mu}+\bar{X}^{\mu})=X^{\mu} + i\Theta\sigma^{\mu}\bar{\Theta} \ec\\
X_{-}^{\mu} &= i(X^{\mu}-\bar{X}^{\mu})= 2\Theta\sigma^{\mu}\bar{\Theta} \ec
\end{align}
which are manifestly odd and even under $z_1 \leftrightarrow z_2$, respectively.

The above constraints are sufficient to completely determine $t(X,\Theta,\bar{\Theta})$ up to an overall numerical factor. In particular, \cite{Fortin:2011nq} found\footnote{We find a slightly different coefficient then \cite{Fortin:2011nq} for the second term in the square brackets of \eqref{eq:tm}.}  two structures corresponding to spin-$\l$ sprimary operators with zero R-charge, which take the form\footnote{Round brackets around Lorentz indices denote symmetrization, which is defined by averaging over all permutations.}
\begin{align}
t^{\mu_1\cdots\mu_\l}_+(X,\Theta,\bar{\Theta}) &\equiv \frac{X_+^{(\mu_1}\cdots X_+^{\mu_\l)}}{(X\cdot\bar{X})^{2-\frac{1}{2}(\Delta-\l)}} \left[ 1 - \frac{1}{4} \left( \Delta - \l - 4 \right) \left( \Delta + \l - 6 \right) \frac{\Theta^2\bar{\Theta}^2}{X\cdot\bar{X}} \right] \ec \label{eq:tp} \\
t^{\mu_1\cdots\mu_\l}_-(X,\Theta,\bar{\Theta}) &= \frac{X_+^{(\mu_1}\cdots X_+^{\mu_{\l-1}}}{(X\cdot\bar{X})^{2-\frac{1}{2}(\Delta-\l)}} \left[ X_-^{\mu_\l)} - \frac{\left( \Delta - \l -4 \right)}{\Delta-2} \frac{\left(X_-\cdot X_+\right) X_+^{\mu_\l)}}{X\cdot\bar{X}} \right] \label{eq:tm} \ed
\end{align}

Under $z_1 \leftrightarrow z_2$ the structures \eqref{eq:tp} and \eqref{eq:tm} transform as
\begin{align}
t^{\mu_1\cdots\mu_\l}_{\pm} \xrightarrow{z_1 \leftrightarrow z_2} \pm(-)^\l t^{\mu_1\cdots\mu_\l}_{\pm} \ed
\end{align}
Therefore if $C_{ab,I}^{(\mathbf{r})}$ is (anti-)symmetric in $a$ and $b$, then in \eqref{JJO1}, the structure $t_+^{\mu_1\cdots\mu_\l}$ appears for (odd) even $\l$ and $t_-^{\mu_1\cdots\mu_\l}$ for (even) odd $\l$.

The $\l=0$ case is special since there is no structure for $\Delta\neq 2$ which is odd under $z_1\leftrightarrow z_2$ (see appendix \ref{scalar}). Therefore in this case only scalar sprimaries in representations which arise from the symmetric product of two adjoints can contribute to \eqref{JJO1} with the structure
\begin{align}
t_+(X,\Theta,\bar{\Theta}) &\equiv \frac{1}{(X\cdot\bar{X})^{2-\frac{1}{2}\Delta}} \left[ 1 - \frac{1}{4} \left( \Delta - 4 \right) \left( \Delta - 6 \right) \frac{\Theta^2\bar{\Theta}^2}{X\cdot\bar{X}} \right] \ed
\end{align}
The $\Delta=2$ scalar in the adjoint representation corresponds to\footnote{There could be other conserved currents in the theory, but those would appear in the singlet representation.} $\cO_I = \cJ_a$. In that case the 3-point function is completely determined (for the canonically normalized current) by the Ward identities to be \cite{Osborn:1998qu}, 
\begin{align}
\langle \cJ^a(z_1) \cJ^b(z_2) \cJ^c(z_3) \rangle &= \frac{1}{x_{\bar{1}3}^2 x_{\bar{3}1}^2 x_{\bar{2}3}^2 x_{\bar{3}2}^2} \left[ i\frac{f^{abc} \tau}{128\pi^6} \left( \frac{1}{X^2} - \frac{1}{\bar{X}^2} \right) + \frac{d_{abc} \kappa}{256\pi^6} \left( \frac{1}{X^2} + \frac{1}{\bar{X}^2} \right) \right] \label{JJJ} \ec
\end{align}
where $\kappa$ is the $\text{Tr}\,\cG^3$ ~`t~Hooft anomaly and $\tau$ is defined through the 2-point function of the canonically normalized current \eqref{jj}.

In addition, we find various contributions to \eqref{JJO1} corresponding to operators which are not in spin-$\l$ Lorentz representations. Those are collected in table \ref{nonSpinl}.

\begin{table}[h]

\begin{tabular}{|c|c|c|c|c|}
\hline

$\left(j,\bar{j}\right)$ & $R$ & $\Delta$  & $z_1\leftrightarrow z_2$ & $t(X,\Theta,\bar{\Theta})$ \\ 

\hline 

$\left(\frac{\l-1}{2},\frac{\l}{2}\right)$ & $1$ & $\l+\frac{7}{2}$ & $(-)^\l$ & $X^{\alpha_1}_{~(\dot{\alpha}_1}\cdots X^{\alpha_{\l-1}}_{~\phantom{(}\dot{\alpha}_{\l-1}}\bar{\Theta}^{\phantom{\alpha}}_{\dot{\alpha}_{\l})}$ \\

\hline

$\left(\frac{1}{2},0\right)$ & $1$ & $\frac{3}{2}$ & $+$ & $(X)^{-4}\cdot X^{\alpha}_{~\dot{\alpha}}\bar{\Theta}^{\dot{\alpha}}$\\

\hline

$\left(\frac{\l}{2},\frac{\l-1}{2}\right)$ & $-1$ & $\l+\frac{7}{2}$ & $(-)^\l$ & $X^{(\alpha_1}_{~~\dot{\alpha}_1}\cdots X^{\phantom{(}\alpha_{\l-1}}_{~~\dot{\alpha}_{\l-1}}\Theta^{\alpha_\l)}_{\phantom{\dot{\alpha}_{\l}}}$ \\

\hline

$\left(0,\frac{1}{2}\right)$ & $-1$ & $\frac{3}{2}$ & $+$ & $(X)^{-4}\cdot X^{\alpha}_{~\dot{\alpha}}\Theta_{\alpha}$ \\

\hline

$\left(\frac{\l+1}{2},\frac{\l-1}{2}\right)$ & $0$ & $\geq \l+3$ & $(-)^\l$ & $(X\cdot\bar{X})^{\frac{\Delta-\l-5}{2}} X^{(\alpha_1}_{+~\dot{\alpha}_1}\cdots X^{\alpha_\l}_{+~\dot{\alpha}_\l}X^{|\dot{\alpha}_\l|\alpha_{\l+1})}_{-}$ \\

\hline

$\left(\frac{\l-1}{2},\frac{\l+1}{2}\right)$ & $0$ & $\geq \l+3$ & $(-)^\l$ & $(X\cdot\bar{X})^{\frac{\Delta-\l-5}{2}} X^{\alpha_1}_{+~(\dot{\alpha}_1}\cdots X^{\alpha_\l}_{+~\dot{\alpha}_\l}X_{-~|\alpha_{\l}|\dot{\alpha}_{\l+1})}$ \\

\hline
\end{tabular}
\caption{Structures corresponding to superconformal primaries in the $\cJ\times\cJ$ OPE, in Lorentz representations with $j\neq\bar{j}$.}
\label{nonSpinl}
\end{table}

Let us discuss some properties of the operators listed in table \ref{nonSpinl}. The $\left(\frac{1}{2},0\right)$ structure in the second entry of table  \ref{nonSpinl} (and its $\left(0,\frac{1}{2}\right)$ conjugate) actually arises from a larger family of  structures \mbox{$t^{\alpha(\l)}_{~\dot{\alpha}(\l-1)} = (X^2)^{-\l-1} X^{(\alpha_1}_{~\dot{\alpha}_1}\cdots X^{\alpha_\l)}_{~\dot{\alpha}_\l}\bar{\Theta}^{\dot{\alpha}_\l}$}, which satisfies all the constraints\footnote{We use the notation: $t^{\alpha(\l),\dot{\alpha}(\l')}\equiv t^{\alpha_1\cdots\alpha_\l,\dot{\alpha}_1\cdots\dot{\alpha}_{\l'}}$.}. These structures correspond to operators with dimension $\Delta=\frac{5}{2} - \l$ which violate the unitarity bound for $\l\neq 1$, $\Delta \ge |\frac{3}{2}R-j+\bar{j}|+j+\bar{j}+2=\l+\frac{5}{2}$. The $\l=1$ structure however, corresponds to a chiral operator ($\bar{Q}_{\dot{\alpha}}\Psi_{\alpha}=0$), in which case the unitarity bound is modified to $\Delta=\frac{3}{2}R\ge j+1$. The corresponding operator saturates this bound, so it is in fact a free chiral fermion.

When the zero R-charge $\left(\frac{\l+1}{2},\frac{\l-1}{2}\right)$ operators saturate the unitarity bound $\Delta\ge \l+3$, they decompose into two short representations as follows:
\begin{align}
\left(\frac{\l + 1}{2},\frac{\l - 1}{2}\right) \xrightarrow{\Delta\rightarrow \l+3} \left(\frac{\l+1}{2},\frac{\l - 1}{2}\right)_{\mathrm{short}} \!\!\oplus \left(\frac{\l}{2},\frac{\l - 1}{2}\right)_{\mathrm{short}}\ed \label{short}
\end{align}
The shortening condition is $Q^{\alpha_1} \cO_{\alpha_1\cdots\alpha_{\l+1}\ec\dot{\alpha}_1\cdots\dot{\alpha}_{\l-1}}=0$. The resulting structure for the  short representation $\left(\frac{\l}{2},\frac{\l - 1}{2}\right)_{\mathrm{short}}$ appears as the third entry of table \ref{nonSpinl}, so these two series of structures are actually related. A similar story holds for the representations $\left(\frac{\l-1}{2},\frac{\l+1}{2}\right)$ and $\left(\frac{\l-1}{2},\frac{\l}{2}\right)$.
This decomposition into short multiplets matches the one described, e.g. on \cite{Gadde:2010en}, which also specifies where the spin $\l$ conformal primaries reside after the decomposition. 

Short representations such as $\left(\frac{\l}{2},\frac{\l - 1}{2}\right)_{\mathrm{short}}$, can certainly appear, at least for free theories. They can be constructed in the following way, using the current $J^a$ as the basic building block 

\begin{align}
\cO^{ab}_{\alpha(\l)\dot{\alpha}(\l-1)} &=    \begin{cases}
        J^{(a}\overleftrightarrow{\partial}_{\!\!\alpha_1\dot{\alpha}_1} \cdots \overleftrightarrow{\partial}_{\!\!\alpha_{\l-1}\dot{\alpha}_{\l-1}}  Q_{\alpha_\l} J^{b)} 
        	-  \left(Q_{\alpha_\l} J\right)^{(a} \overleftrightarrow{\partial}_{\!\!\alpha_1\dot{\alpha}_1} \cdots \overleftrightarrow{\partial}_{\!\!\alpha_{\l-1}\dot{\alpha}_{\l-1}} J^{b)}      
        	       & \l=\text{even} \ec \\
        J^{[a} \overleftrightarrow{\partial}_{\!\!\alpha_1\dot{\alpha}_1} \cdots \overleftrightarrow{\partial}_{\!\!\alpha_{\l-1}\dot{\alpha}_{\l-1}} 
        Q_{\alpha_\l} J^{b]} 
            - \left(Q_{\alpha_\l} J\right)^{[a} \overleftrightarrow{\partial}_{\!\!\alpha_1\dot{\alpha}_1} \cdots \overleftrightarrow{\partial}_{\!\!\alpha_{\l-1}\dot{\alpha}_{\l-1}} J^{b]}& \l=\text{odd} \ed
    \end{cases} 
\label{short_ops}
\end{align}
Symmetrization over Lorentz is to be understood. One can verify that these are superconformal primaries and satisfy the shortening condition $Q^{\alpha_1} \cO^{ab}_{\alpha_1 \cdots \alpha_{\l}\ec\dot{\alpha}_1 \cdots \dot{\alpha}_{\l-1}}=0$, by using the superconformal algebra and the fact that $Q^2 J^a(x) = \bar{Q}^2 J^a(x)=0$.


In the following section we will describe the application of these results to the conformal block decomposition of the 4-point function of $J^a(x)$.

\subsection{Superconformal Blocks}
The structures for the 3-point function $\langle \cJ_a(z_1) \cJ_b(z_2) \cO^i_I(z_3) \rangle$, found in the previous section, relate the $\cJ^a\times\cJ^b$ OPE coefficients of primary super-descendants of $\cO^i_I$, to the coefficient of the superconformal primary. 
The sum over primary operators in the conformal block decomposition of the current 4-point function, can then be rearranged as a sum over superconformal primary operators, with ``superconformal blocks'' replacing the usual conformal blocks. The superconformal blocks are linear combinations of the usual conformal blocks, which take into account the relations between OPE coefficients of the primary operators in each super-multiplet.

For the purposes of this paper, we are interested in these  relations for the $J^a\times J^b$ OPE. These can be obtained by setting $\theta_{1,2}=\bar{\theta}_{1,2}=0$ in the various expressions for \eqref{JJO1}, expanding in $\theta_3$ and $\bar{\theta}_3$ and disentangling the various primary super-descendants in this expansion.

The superconformal blocks for spin-$\l$ sprimaries, corresponding to the $t_+$ and $t_-$ structures in equations \eqref{eq:tp} and \eqref{eq:tm}, were computed in \cite{Fortin:2011nq}. The result is\footnote{Equation \eqref{Gm} fixes a mistake in the superconformal block which was found in \cite{Fortin:2011nq}. We are grateful to J.F~Fortin, K.~Intriligator and A.~Stergiou for discussions on this point.} \footnote{We are grateful to Z.U.~Khandker, D.~Li, D.~Poland and D.~Simmons-Duffin for pointing out a mistake in (\ref{Gp}) in an earlier version of this paper. Their full analysis can be found in \cite{KLPSD}.}.
\begin{align}
\cG^+_{\Delta,\l}(u,v) &= g_{\Delta,\l}(u,v) + \frac{ (\Delta-2)^2 (\Delta + \l) (\Delta - \l - 2) }{ 16 \Delta^2 (\Delta + \l + 1) (\Delta - \l - 1) } g_{\Delta+2,\l}(u,v) \ec \label{Gp}\\
\cG^-_{\Delta,\l}(u,v) &= g_{\Delta+1,\l+1}(u,v)+
\frac{(\l+2)^2 (\Delta+\l+1)(\Delta-\l-2)}{\l^2 (\Delta -\l-1) (\Delta + \l)} g_{\Delta+1,\l-1}(u,v) \ed \label{Gm}
\end{align}

Depending on the spin and global symmetry representation of the spin-$\l$ sprimary, either  $\cG^+_{\Delta,\l}$ or $\cG^-_{\Delta,\l}$ should be used in the superconformal block decomposition. In particular, define
\begin{align}
\cG^{(\mathbf{r})_s}_{\Delta,\l} &=
    \begin{cases}
        \cG^+_{\Delta,\l}\ec & \l=\text{even}  \\
        \cG^-_{\Delta,\l}\ec & \l=\text{odd}
    \end{cases} \ec \\
\cG^{(\mathbf{r})_a}_{\Delta,\l} &=
\begin{cases}
        \cG^+_{\Delta,\l}\ec & \l=\text{odd}  \\
        \cG^-_{\Delta,\l}\ec & \l=\text{even}
    \end{cases} \ed
\end{align}
If the representation of the sprimary is in the (anti-)symmetric product then one should use ($\cG^{(\mathbf{r})_a}_{\Delta,\l}$) $\cG^{(\mathbf{r})_s}_{\Delta,\l}$.

For the operators corresponding to the structures in table \ref{nonSpinl} there is only one primary super-descendant which can contribute to the $J^a\times J^b$ OPE (i.e. which is in a spin-$\l$ Lorentz representation). Therefore, there are no special relations between OPE coefficients inside each multiplet in those cases. 
In particular, the $\left(\frac{\l\pm 1}{2},\frac{\l\mp 1}{2}\right)$ sprimaries contain a spin-$\l$ primary super-descendant of dimension $\Delta\ge  \l+4$, which is obtained by acting on the sprimary with $Q_{\alpha}$ and $\bar{Q}_{\dot{\alpha}}$ appropriately\footnote{When the unitarity bound is saturated the multiplet decomposes to two short multiplets as in \eqref{short}. The above spin-$\l$ primary superdescendant sits in the second factor on the RHS of \eqref{short} as can be seen e.g., in \cite{Gadde:2010en}.}. 

To summarize, we can write the $s$-channel decomposition as follows:
\begin{gather}
\langle J^m_{~i}(x_1) J^n_{~j}(x_2) J^p_{~k}(x_3) J^q_{~l}(x_4) \rangle_{s} = \frac{1}{x_{12}^{4}x_{34}^{4}} \Bigg\{  \notag\\
\sum_{\mathbf{r} \in Adj. \otimes_s Adj.} \id{mn}{ij}{pq}{kl}{\mathbf{r}} \biggl[ \,\,\sum_{\substack{ \cO_{\alpha(\l),\dot{\alpha}(\l)} \\ \Delta\ge \l+2 }} \!\!\!\! p_{\cO}\, \cG_{\Delta,\l}^{(\mathbf{r})_s}(u,v) \,\,\,+\!\! \sum_{\substack{ \cO_{\alpha(\l\pm 1),\dot{\alpha}(\l\mp 1)} \\ \Delta\ge \l+3 }} \!\!\!\!\!\! p_{\cO}\, g_{\Delta+1,\l}(u,v)\biggr] \notag\\ 
+ \sum_{\mathbf{r} \in Adj. \otimes_a Adj.} \id{mn}{ij}{pq}{kl}{\mathbf{r}} \biggl[\,\,\sum_{\substack{ \cO_{\alpha(\l),\dot{\alpha}(\l)} \\ \Delta_{\cO}\ge \l+2 }} \!\!\!\! p_{\cO} \, \cG_{\Delta,\l}^{(\mathbf{r})_a}(u,v) \,\,\,+\!\! \sum_{\substack{ \cO_{\alpha(\l\pm 1),\dot{\alpha}(\l\mp 1)} \\ \Delta\ge \l+3 }} \!\!\!\!\!\! p_{\cO} \, g_{\Delta+1,\l}(u,v)\biggr]\Bigg\} \ec\label{Neq1SR}
\end{gather}
where we separated the sum over representations to sums over symmetric and anti-symmetric representations. The sums in the square brackets are over sprimary operators in the $J^a\times J^b$ OPE in the indicated Lorentz representation\footnote{In the sum over operators in the Lorentz reprsentations $\left(\frac{\l\pm 1}{2},\frac{\l\mp 1}{2}\right)$, we implicitly include also the short $\left(\frac{\l}{2}\ec\frac{\l-1}{2}\right)$ and $\left(\frac{\l-1}{2}\ec\frac{\l}{2}\right)$ operators.}. Note that for the operators in table \ref{nonSpinl}\footnote{For spin-$\l$ sprimaries the summation is over even and odd spins regardless of whether the representation of $\cO$ is in the symmetric or anti-symmetric product of the current operators. This is because an even (odd) spin sprimary contains odd (even) super-descendant conformal primaries.} only (odd) even $\l$ appears for (anti-)symmetric representations. A similar expression holds for the $t$-channel. The final result for the sum-rules in the $\cN=1$ case is obtained using the appropriate conformal or superconformal blocks in the adjoint scalar sum-rules written in section \ref{adjScalar}.

We wrote the coefficient $p_{\cO}$ in \eqref{Neq1SR} with some abuse of notation to avoid clutter. It should be understood that it denotes the coefficient which was defined in \eqref{SUnSR} for the  appropriate operator. In particular, for a spin-$\l$ sprimary it denotes the coefficient of the sprimary if $\cG^+_{\Delta,\l}$ is used, while if $\cG^-_{\Delta,\l}$ is used, it denotes the coefficient of the spin-$\l\!+\!1$ super-descendant. For the sprimaries in table \ref{nonSpinl} it denotes the coefficient of the spin-$\l$ super-descendant.

A check of the above results can be obtained by decomposing the $\mathcal{N}=2$ superconformal blocks found in \cite{Dolan:2001tt}. This was described in some detail \cite{Fortin:2011nq}, though not carried out explicitly. We verified that this decomposition is consistent only if we include the operators in Table \ref{nonSpinl}.

\section{Bounds on Current Central Charges}
\label{results}
Having written down the sum-rules, including the SUSY constraints, we are now ready to apply any of the methods developed in
\cite{Rattazzi:2008pe,Poland:2010wg, Rattazzi:2010gj,Poland2011}
to find bounds on OPE coefficients. The basic strategy for obtaining 
such bounds involves converting the problem into a system of constraints 
for every possible operator in the spectrum and is reviewed in Appendix \ref{bootstrap_bounds}.

In SCFTs, the $\Delta=2$ adjoint scalar sits at the top of the current 
supermultiplet. Thus, one can effectively use it to place a bound applicable for
every $\cN=1$ theory with $SU(N)$ global symmetry.
Specifically, we have obtained an lower bound on $\tau$, 
the coefficient of the current two-point function.

\subsection{Bounds on OPE coefficients in $U(1)$ SCFTs}
The leading terms in the $JJ$ OPE, when $J$ is canonically normalized, take the form
\begin{align}
J(x) J(0)=\tau \frac{\mathds{1}}{16\pi^4x^4}+\frac{\kappa}{\tau} \frac{J(0)}{16\pi^2x^2} +c^i\frac{\cO_i(0)}{x^{4-\Delta _i}}+\cdots.
\end{align}
We first attempt to obtain a bound for the OPE coefficient $\lambda_J$. In our normalization \eqref{2pnt}, \eqref{3pnt}, this is nothing but $\lambda_J = \frac{\kappa}{4 \tau^{3/2}}$.
In this case, which has fewer sum rules, we used the procedure and parameters described in \cite{Poland2011} with $k=10$, 
and obtained an upper bound for the OPE coefficient $|\lambda_J| < 5.38$.\footnote{We are grateful to D.~Li for pointing out to us a mistake in the $\lambda_J$ bound that appeared in the previous version of this paper.}

\subsection{Bounds on OPE Coefficients and on $\tau$ in $SU(N)$ SCFTs}
The major difference from the $U(1)$ case arises from the fact that there are now several different tensor structures appearing in the OPE:
three in the case of $SU(2)$, five in the case of $SU(3)$ and seven in the generic case $SU(N)$ for $N>3$.
As shown, for example, in \cite{Poland:2010wg,Rattazzi:2010yc,Vichi:2011ux, Poland2011},
one can use a vectorial linear functional in order to obtain a bound when several sum-rules are involved.


In the non-abelian case, the $JJ$ OPE for the canonically normalized current takes the form
\begin{align}
\label{JJOPE}
J_a(x) J_b(0) &= \tau \frac{\delta _{ab}\mathds{1}}{16\pi ^4x^4}+\frac{k d_{abc}}{\tau} \frac{J_c(0)}{16\pi ^2x^2}+f_{abc} \frac{x^\mu j_\mu ^c (0)}{24\pi ^2 x^2}+c_{ab}^i\frac{\cO_i(0)}{x^{4-\Delta _i}}+\cdots \ec \\
j_{\mu}^a(x) j^b_{\nu}(0) &= 3\tau\delta^{ab} \frac{ I_{\mu\nu}(x)}{4\pi^4x^6} \mathds{1} +\cdots \ed
\end{align}

\begin{figure}[h]
    \includegraphics[width=1 \linewidth,scale=0.45]{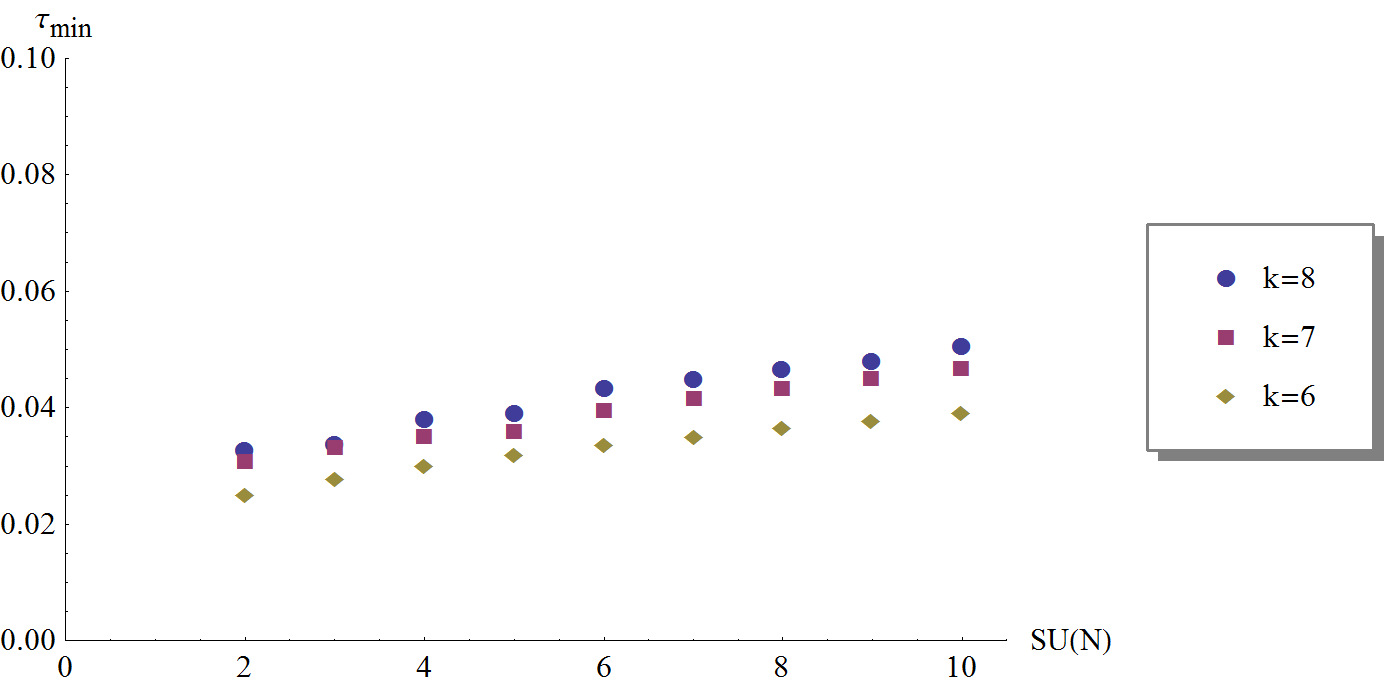}
    \centering
    \caption{Lower bounds on $\tau$ for different gauge groups $SU(N)$, obtained for different search space sizes $k$. }     
    \label{bounds_fig}
\end{figure}

We want to place a bound on $\tau$. This can be done by isolating the contribution of $j^a_{\mu}$ in the sum-rules and placing a bound on its OPE coefficient. In our normalization \eqref{2pnt}, \eqref{3pnt} we have
\begin{align}
\lambda_{ab,c}^{j_{\mu}} = \frac{1}{\sqrt{3 \tau}} f_{abc} \ed
\end{align}
Therefore, $|\lambda_{j_{\mu}}|=\frac{1}{\sqrt{3 \tau}}$ in \eqref{OPEcf}. The OPE coefficient enters the sum-rule as the coefficient $p_{j_{\mu}}$ of the conformal block $g_{3,1}$ (see \eqref{SUnSR})\footnote{This corresponds to the structure \eqref{JJJ}. In this channel there is only one operator in the $JJ$ OPE and therefore no superconformal block.}, and due to our normalization of the conformal blocks chosen without the $(2)^{-\l}$ factor we have,  
\begin{align}
\label{tau_bound}
p_{j_{\mu}} = \frac{1}{2} |\lambda_{j_{\mu}}|^2 = \frac{1}{6\tau} \ed
\end{align}
We can obtain an upper bound on $p_{j_\mu}$, which translates into a lower bound on $\tau$. Figure \ref{bounds_fig} shows the lower bounds on $\tau$ obtained for different values of the gauge group size $N$. Due to numerical difficulties, the bounds for $k > 6$ is not optimal, but it does satisfy the constraints.

Note that the bound increases with $N$, as one would expect. Indeed, we can think of $SU(2)$ as a subgroup of $SU(N)$ with $N>2$. In that case the generators of $SU(N)$, which are not part of the $SU(2)$ subgroup, would appear in the singlet representation of the $SU(2)$ current-current OPE. Thus, for consistency, the bound for $SU(2)$ must be weaker than the bound for $N>2$. This is indeed the case. 

\begin{figure}[h]
    \includegraphics[width=1 \linewidth,scale=0.45]{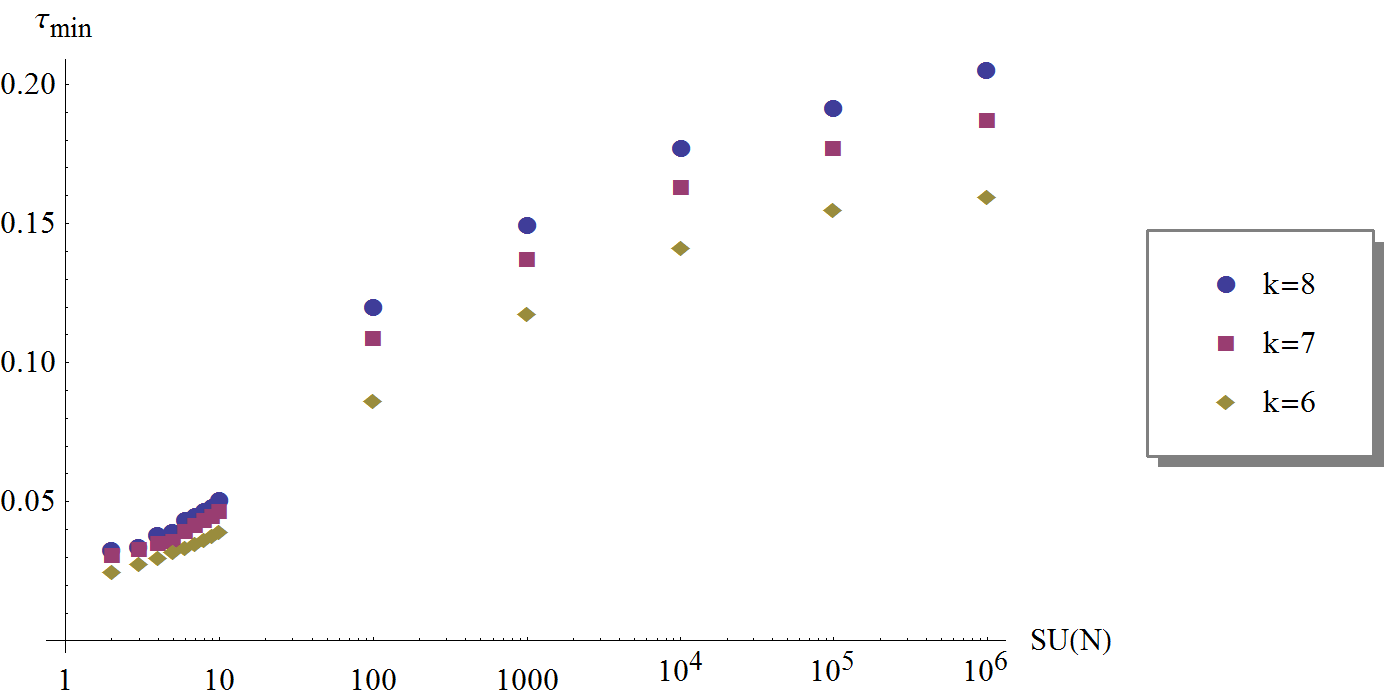}
    \centering
    \caption{Lower bounds on $\tau$ for different gauge groups $SU(N)$ with high values of $N$, obtained for different search space sizes $k$. }     
    \label{high_bounds_fig}
\end{figure}

In Figure \ref{high_bounds_fig} we show the bounds for (some) very large values of $N$ as well.
The bound again rises with $N$, though very slowly. 


These results are consistent with the results of \cite{Poland:2010wg,Poland2011}, which were obtained by analyzing the 4-point function of a chiral field $\Phi$ in the fundamental representation of $SU(N)$. In those works the bounds on $\tau$ were obtained as a function of $\Delta_{\Phi}$. For $\Delta_{\Phi}=1$ one finds $\tau\ge 1$. The saturation of this bound corresponds to the free theory value of $\tau$. Eventually, for high enough dimensions of $\Phi$ the bound drops well below $1$. Since we do not assume any particular field content, one would expect our bounds to be weaker then the ones found in \cite{Poland:2010wg,Poland2011} for the entire range of $\Delta_{\Phi}$, and this is indeed the case. 

Nevertheless, it is still slightly puzzling that the bounds we find appear to be much weaker than those of \cite{Poland:2010wg,Poland2011}. While this could be simply due to the fact that we make less assumptions on the theory, the following argument offers an alternative explanation.



Consider the free theory of one fundamental chiral field. In this theory the $SU(N)$ currents $j_{B\mu}^a$, $j_{F\mu}^a$ constructed from the boson and fermion fields are separately conserved, while only the combination $j_{\mu}^a=j_{B\mu}^a+j_{F\mu}^a$  sits in a current multiplet (and the current central charge corresponding to $j_{\mu}^a$ is $\tau=1$). The additional symmetries contaminate the OPE coefficient of $g_{3,1}$ out of which we have been extracting our bounds on $\tau$. In fact, the 4-point function of $J^a$ in the free $\cN=1$ theory is obviously the same as in the non-supersymmetric theory of one scalar field. Therefore, if we interpret the coefficient of $g_{3,1}$ as $1/6\tau$ we would obtain $\tau=1/3$; the value of $\tau$ for one free fundamental scalar. As a result, using our method we cannot expect our bounds to be stronger then $\tau\ge 1/3$.   

As we just explained our interpretation of the coefficient of $g_{3,1}$ is incorrect in the free theory, since this coefficient receives other contributions.  These additional contributions are special to the free theory, as the corresponding conserved currents reside in multiplets which contain conserved higher-spin currents \cite{Maldacena:2011jn}. It would be interesting to remove the free theory from the numerical analysis, by introducing small gaps in the dimensions to exclude conserved higher-spin currents from the $JJ$ OPE.

In fact, we expect this to improve the bounds dramatically, especially for large values of $N$. For instance, it was shown in \cite{Beem:2013sza} that in interacting $\cN=2$ CFTs $\tau \ge N$ for $N \ge 3$. In those cases there is therefore a discontinuous jump in the bound compared to free theories which have $\tau = O(N^0)$. We find it plausible that similar results hold also for $\cN=1$ CFTs. It would be interesting to study the interplay between the size of the dimension gaps needed to see a jump in the bound and the size of $N$. We leave this to future work.
As a preliminary result, we found that the bound is $\tau > 3.82$ for $SU(10000)$ with $k=6$ when one assumes a gap of $0.1$ above the 
unitarity bound for all spin $\ell$ operators which are not shortened due to the global symmetry. 


Finally, if the theory has a gravity dual, then in our normalization we have \cite{Freedman:1998tz}
\begin{equation}
\tau = 8 \pi^2 \frac{R_{AdS}}{g^2},
\end{equation}
where $g$ is the coupling constant of the non-abelian gauge theory in the bulk, which matches the $SU(N)$ global symmetry.
Thus, one can obtain an upper bound on $g^2/R_{AdS}$, meaning that the gauge coupling cannot become arbitrarily large in the bulk theory. 
This argument has been used in \cite{Poland:2010wg} to claim
that such a bound exists in a bulk theory in the presence of a charged scalar. Here we see that it exists regardless of the type of excitation, and it is just a consequence of having a holographic dual. 
It would be very interesting to understand why such a bound exists from the bulk perspective. As an order of magnitude estimation,
a lower bound of $\tau \ge 1$ at $N\rightarrow \infty$ translates to $g^2/R_{AdS} \le 8 \pi^2$.

\subsection*{Acknowledgments}
\label{s:acks}
We are grateful to O.~Aharony for many suggestions, enlightening conversations and collaboration at early stages of this project. We would also like to thank D.~Simmons-Duffin and B.C. van Rees for discussions and helpful suggestions. We are especially grateful to J.F.~Fortin, K.~Intriligator and A.~Stergiou for useful comments on an early version of this manuscript. We would like to thank S.~Rychkov for very helpful comments on the first version of this paper, which have helped us pinpoint a mistake in the overall normalization of $\tau$. Finally, we are grateful to Z.U.~Khandker, D.~Li, D.~Poland and D.~Simmons-Duffin for sharing their results with us, thereby correcting a mistake in the earliar version of this draft. 
This work was supported in part by an Israel Science Foundation (ISF) center of excellence grant, by the German-Israeli Foundation for Scientific Research and Development and by the Minerva Foundation.


\appendix

\ytableausetup{mathmode, boxsize=1.7em,aligntableaux=center}
\section{Product of Two $SU(N)$ Adjoints}\label{group}

Let us decompose the tensor product of two $SU(N)$ adjoints into irreducible representations. Generally this decomposition contains $7$ irreducible representations:
\begin{center}
\scalebox{0.6}{
\begin{ytableau}
1  &  \\
2  \\
\none[\vdots] \\
\scriptstyle N-1
\end{ytableau}
}
$\otimes$
\scalebox{0.6}{
\begin{ytableau}
1  &  \\
2  \\
\none[\vdots] \\
\scriptstyle N-1
\end{ytableau}
}
=
\begin{tabular}{llllllll}
& & & & & & &\\
\scalebox{0.6}{
\begin{ytableau}
1 &  &  &  \\
2 & \\
\none[\vdots] & \none[\vdots] \\
\scriptstyle N-1 &
\end{ytableau}
}
$\oplus$ &
\scalebox{0.6}{
\begin{ytableau}
1 & \\
2 & \\
\none[\vdots] \\
\scriptstyle N-2
\end{ytableau}
}
$\oplus$ &
\scalebox{0.6}{
\begin{ytableau}
1 & & \\
2 \\
\none[\vdots] \\
\scriptstyle N-2
\end{ytableau}
}
$\oplus$ &
\scalebox{0.6}{
\begin{ytableau}
1 & & \\
2 & & \\
\none[\vdots] & \none[\vdots]\\
\scriptstyle N-1 & \scriptstyle N-1
\end{ytableau}
}
$\oplus$ &
\scalebox{0.6}{
\begin{ytableau}
1  &  \\
2  \\
\none[\vdots] \\
\scriptstyle N-1
\end{ytableau}
}
$\oplus$ &
\scalebox{0.6}{
\begin{ytableau}
1  &  \\
2  \\
\none[\vdots] \\
\scriptstyle N-1
\end{ytableau}
}
$\oplus$ &
1
\vspace{3mm}
\\

$(S,\bar{S})_s$ & $(A,\bar{A})_s$ & $(S,\bar{A})_a$ & $(A,\bar{S})_a$ & $Adj_s$ & $Adj_a$ & $1_s$
\end{tabular}
\end{center}

For example, the notation $(A,\bar{S})_a$ means that the (traceless) tensor carrying the representation is anti-symmetric in the two fundamental indices and symmetric in the anti-fundamental indices. The subscript ($a$) $s$ means that this representation is in the (anti-)symmetric product of the two adjoints. Note that $(S,\bar{A})_a$ and $(A,\bar{S})_a$ are complex conjugate representations. The dimensions of the less familiar representations are
\begin{align}
\left|(S,\bar{S})_s\right| &= \frac{(N+3)N^2(N-1)}{4} \ec \\
\left|(A,\bar{A})_s\right| &= \frac{(N+1)N^2(N-3)}{4} \ec \\
\left|(S,\bar{A})_a\right| &= \left|(A,\bar{S})_a\right| = \frac{(N+2)(N+1)(N-1)(N-2)}{4} \ed
\end{align}

These formulas can be checked up to $N=8$ in the tables of \cite{Slansky:1981yr}. Note that $SU(2)$ and $SU(3)$ are special cases. $(A,\bar{A})_s$ does not exist in either $SU(2)$ or $SU(3)$ and the conjugate pair $(S,\bar{A})_a$, $(A,\bar{S})_a$ do not exist in $SU(2)$. In addition the adjoint representation in the product of two $SU(2)$ adjoints comes only from the anti-symmetric combination.

\subsection{Identity Matrices}

Let us determine the identity matrices $(\mathds{1}_{\mathbf{r}})_{ab,cd}$ defined around \eqref{SUnSR}. In particular we will write those matrices in the fundamental representation basis,
\begin{align}
\id{mn}{ij}{pq}{kl}{\mathbf{r}}\equiv (\mathds{1}_{\mathbf{r}})_{ab,cd} (T^a)^m_{~i} (T^b)^n_{~j} (T^c)^p_{~k} (T^d)^q_{~l} \ed
\end{align}
This is more convenient since the symmetry properties of the representations $\mathbf{r}$ in the tensor product are most easily expressed in the fundamental basis. 

The identity matrix can be constructed by symmetrizing and removing traces appropriately from the tensor $\delta^p_i\delta^q_j\delta^m_k\delta^n_l$. This determines the identity matrices up to an overall normalization. The overall sign of the identity matrices is determined by reflection-positivity as described in \cite{Rattazzi:2010yc}.

Up to an overall positive normalization (which can be absorbed in the OPE coefficients) we find\footnote{The adjoint representations appear in the OPE as $J^a \times J^b \sim f^{abc}\cO_c + d^{abc}\cO'_c$. The expressions for the identities in fundamental representation indices where obtained from $\id{mn}{ij}{pq}{kl}{(Adj.)_s} \equiv d^{abe}d^{dce} (T^a)^m_i(T^b)^n_j(T^c)^p_k(T^d)^q_l$, and $\id{mn}{ij}{pq}{kl}{(Adj.)_a} \equiv f^{abe}f^{dce} (T^a)^m_i(T^b)^n_j(T^c)^p_k(T^d)^q_l$.}
\begin{align}
\id{mn}{ij}{pq}{kl}{(S,\bar{S})_s} &= \delta_{(i}^p \delta_{j)}^q \delta_{(k}^m \delta_{l)}^n -\frac{1}{N+2}\left[ \delta_{(i}^{(m|} \delta_{j)}^{(p} \delta_{(k}^{q)} \delta_{l)}^{|n)} -\frac{2}{N+1} \delta_{(i}^m \delta_{j)}^n \delta_{(k}^p \delta_{l)}^q \right] \ec \notag\\
\id{mn}{ij}{pq}{kl}{(A,\bar{A})_s} &= \delta_{[i}^p \delta_{j]}^q \delta_{[k}^m \delta_{l]}^n -\frac{1}{N-2}\left[ \delta_{[i}^{[m|} \delta_{j]}^{[q} \delta_{[k}^{p]} \delta_{l]}^{|n]} + \frac{2}{N-1} \delta_{[i}^m \delta_{j]}^n \delta_{[k}^q \delta_{l]}^p \right] \ec \notag\\
\id{mn}{ij}{pq}{kl}{(S,\bar{A})_a \times (A,\bar{S})_a} &= \delta_{(i}^p \delta_{j)}^q \delta_{[l}^m \delta_{k]}^n -\frac{1}{N} \delta_{(i}^{[m|} \delta_{j)}^{(q} \delta_{[l}^{p)} \delta_{k]}^{|n]}  \ec \notag\\
\id{mn}{ij}{pq}{kl}{(A,\bar{S})_a \times (S,\bar{A})_a} &= \delta_{[i}^q \delta_{j]}^p \delta_{(k}^m \delta_{l)}^n -\frac{1}{N} \delta_{[j}^{(m|} \delta_{i]}^{[q} \delta_{(k}^{p]} \delta_{l)}^{|n)}  \ec \label{id}\\
\id{mn}{ij}{pq}{kl}{(Adj.)_a} &= \delta_i^n \delta_j^p \delta_k^q \delta_l^m - \delta_i^n \delta_j^q \delta_k^m \delta_l^p - \delta_i^p \delta_j^m \delta_k^q \delta_l^n + \delta_i^q \delta_j^m \delta_k^n \delta_l^p \ec \notag\\
\id{mn}{ij}{pq}{kl}{(Adj.)_s} &= \delta_i^n \delta_j^p \delta_k^q \delta_l^m + \delta_i^n \delta_j^q \delta_k^m \delta_l^p + \delta_i^p \delta_j^m \delta_k^q \delta_l^n + \delta_i^q \delta_j^m \delta_k^n \delta_l^p - \nonumber \notag\\
& \frac{2}{N} \left[ \delta_i^m \delta_j^p \delta_k^q \delta_l^n + \delta_i^m \delta_j^q \delta_k^n \delta_l^p + 2\delta_i^n \delta_j^m \delta_k^q \delta_l^p + \delta_i^n \delta_j^p \delta_k^m \delta_l^q + \delta_i^n \delta_j^q \delta_k^p \delta_l^m  + \right. \nonumber \notag\\
& \left. \delta_i^p \delta_j^m \delta_k^n \delta_l^q + \delta_i^p \delta_j^n \delta_k^q \delta_l^m + \delta_i^q \delta_j^m \delta_k^p \delta_l^n + \delta_i^q \delta_j^n \delta_k^m \delta_l^p \right] + \nonumber \notag\\
& \frac{4}{N^2} \left[ 2\delta_i^m \delta_j^n \delta_k^q \delta_l^p + \delta_i^m \delta_j^p \delta_k^n \delta_l^q +\delta_i^m \delta_j^q \delta_k^p \delta_l^n + 2\delta_i^n \delta_j^m \delta_k^p \delta_l^q + \delta_i^p \delta_j^n \delta_k^m \delta_l^q + \right. \nonumber \notag\\
&\left. \delta_i^q \delta_j^n \delta_k^p \delta_l^m \right] - \frac{16}{N^3}\delta_i^m\delta_j^n\delta_k^p\delta_l^q \ec \notag\\
\id{mn}{ij}{pq}{kl}{(\mathds{1})_s} &= \left( \delta_i^n \delta_j^m - \frac{1}{N}\delta_i^m\delta_j^n \right)\left( \delta_k^q \delta_l^p - \frac{1}{N}\delta_k^p\delta_l^q \right) \notag\ec
\end{align}
where $A_{(i}B_{j)}=A_iB_j+A_jB_i$ and $A_{[i}B_{j]}=A_iB_j-A_jB_i$.

Note that all the representations in the product of two adjoints are real except for $(S,\bar{A})_a$ and $(A,\bar{S})_a$ which are complex conjugates. In a case where the representation is complex the conformal block decomposition depends only on the combination ($s$-channel) $\lambda_{ab,I}\lambda_{cd,\bar{I}} +\lambda_{cd,I}\lambda_{ab,\bar{I}} = |\lambda_{\cO}|^2 \delta^{I\bar{J}} ( C_{ab,I}^{(\mathbf{r})}\bar{C}_{cd,\bar{J}}^{(\bar{\mathbf{r}})} +  C_{cd,I}^{(\mathbf{r})}\bar{C}_{ab,\bar{J}}^{(\bar{\mathbf{r}})})$, so the sum-rule only depends on the combination $\id{mn}{ij}{pq}{kl}{(S,\bar{A})_a} \equiv \id{mn}{ij}{pq}{kl}{(A,\bar{S})_a \times (S,\bar{A})_a}+\id{mn}{ij}{pq}{kl}{(S,\bar{A})_a \times (A,\bar{S})_a }$. This just reflects the fact that since the scalars in the $4$-point function are real, the complex irreducible representations in the OPE must combine to the reducible real sum $\mathbf{r}+\bar{\mathbf{r}}$.

We have verified that these projection operators can be used to decompose both the four point function of a single field in the adjoint and the four point function of an adjoint bilinear in the theory of a single field in the fundamental. These are non-trivial checks as, for $N>4$ they include $24$  equations (for the 24 independent tensors which are products of 4 $\delta$'s) for the $6$ unit projection tensors.

\section{Scalar has no odd structure}
\label{scalar}
We show that there is no $R=0$ sprimary scalar of dimension $\Delta\neq 2$, in the anti-symmetric part of the $J^a\times J^b$ OPE.

The $3$-point function \eqref{JJO1} for such a scalar should satisfy the constraints,
\begin{align}
t(\lambda\bar{\lambda} X, \lambda\Theta,\bar{\lambda}\bar{\Theta}) &= \lambda^{\Delta-4}\bar{\lambda}^{\Delta-4} t(X,\Theta,\bar{\Theta}) \ec \\
t(X,\Theta,\bar{\Theta})|_{z_1\leftrightarrow z_2} &= -t(X,\Theta,\bar{\Theta}) \ec \\
\cD^2t &= \bar{\cD}^2t = 0 \ed
\end{align}
By expanding $t$ in the grassmann variables and using the constraints it is easy to see that
\begin{align}
t(X,\Theta,\bar{\Theta}) = t_0(X) + t_{\mu}(X) \Theta\sigma^{\mu}\bar{\Theta} \ed
\end{align}
Now, using the anti-symmetry in $z_1\leftrightarrow z_2$ (under which $X \to -\bar{X}$) we obtain the equations
\begin{align}
\partial_{\mu} t_0(X) &= \frac{i}{2} \left(t_{\mu}(X) + t_{\mu}(-X)\right) \ec \label{diff1} \\
\Box t_0(X) &= i\partial^{\mu}t_{\mu}(X) \ed
\label{diff2}
\end{align}
In addition, we have the scaling constraints which follow from the ones for $t(X)$
\begin{align}
t_0(\lambda X) &= \lambda^{\Delta-4}t_0(X) \ec \\
t_{\mu}(\lambda X) &= \lambda^{\Delta-5}t_{\mu}(X) \ed
\end{align}

Solving the above scaling constraints in terms of polynomials in $X^{\mu}$ and plugging in the differential equations \eqref{diff1},\eqref{diff2} we see that there is no solution unless $\Delta=2$.

\section{The Numerical Bootstrap}
\subsection{Obtaining Numerical Bounds on OPE coefficients}
\label{bootstrap_bounds}
We now review, briefly, how one obtains upper bounds on OPE coefficients.
To find a bound for the OPE coefficient of a superconformal primary $\cO_0$, with conformal dimension $\Delta_0$ and spin $\ell_0$,
we first isolate it from the sum in \eqref{sum_rule_u1},
moving all other operators to the RHS, obtaining
\begin{equation}
p_{\Delta_0,\ell_0} F_{\Delta_0, \ell_0} = 1 -
\sum_{
\substack{
  \cO_{\Delta, \ell} \in J\times J \\
  \cO_{\Delta, \ell} \neq \mathds{1},\cO_0
}}
p_{\Delta,\ell} F_{\Delta, \ell}(z,\bar{z}) \ed
\end{equation}
Next, we apply a linear functional $\alpha: f(z,\bar{z}) \rightarrow \mathbb{R} $ to the functions $F_{\Delta,\ell}(z,\bar{z})$, demanding that the following conditions hold:
\begin{align}
\label{conditions}
&\alpha\left[F_{\Delta_0, \ell_0}\right] = 1 \ec\\
&\alpha\left[F_{\Delta, \ell}\right] \geq 0\textrm{ for all other operators satisfying the unitarity bounds}\ed
\end{align}
For all linear functionals satisfying these constraints we have
\begin{equation}
p_{\Delta_0,\ell_0} = \alpha[1] - \sum_{\substack{\cO_{\Delta, \ell} \in J\times J \\ \cO_{\Delta, \ell} \neq \mathds{1},\cO_0 }} p_{\Delta,\ell} \alpha\left[F_{\Delta, \ell}\right] \leq  \alpha[1].
\end{equation}
For the last inequality we have used the fact that both the OPE coefficients squared $p_{\Delta,\ell}$ and the functionals applied to the functions $F$ are positive. Then, minimizing $\alpha[1]$ over all functionals satisfying the constraints in \eqref{conditions} can yield an upper bound on $|\lambda_\cO|^2$, if any such functionals can be found.

In four dimensions, lacking analytical tools to solve the infinite-dimensional minimization problem, one is forced to perform this minimization while limiting the search space to a finite dimensional subset of all possible functionals. This procedure yields a valid, though not necessarily tight, bound.
Previous works have found it useful, due to the special properties of the functions
$F_{\Delta, \ell}(z,\bar{z})$,
to use the following test functionals
\begin{equation}
\alpha[F_{\Delta,\ell}(z,\bar{z})] =
\sum_{\substack{m+n \leq 2k \\  m < n}} a_{m n}
\partial_z^m \partial_{\bar{z}}^n 
q(z,\bar{z}) F_{\Delta,\ell}(z,\bar{z}) \Big|_{z=\bar{z}=1/2}.
\label{test_functionals}
\end{equation}
Here, $q(z,\bar{z})$ is some function which does not depend on $\Delta$ or $\ell$
and $k$ is a positive integer limiting the size of the search space. More details can be found in Appendix A of \cite{Poland2011}.
The functions $F_{\Delta, \ell}$ are symmetric with respect to $z \leftrightarrow \bar{z}$, and we must also have $m+n$ even.
The minimization is then over all possible values of $a_{m n}$.

When there is a global symmetry \cite{Vichi:2011ux} there are several such sum-rules. The test functionals 
now take the form
\begin{equation}
\alpha[\vec{V}_{\mathbf{r},\Delta,\ell}(z,\bar{z})] =
\sum_{j=1}^{N_{SR}} \sum_{\substack{m+n \leq 2k \\  m < n}} a_{m n}^j \partial_z^m \partial_{\bar{z}}^n
q(z,\bar{z}) V^{j}_{\mathbf{r},\Delta,\ell}(z,\bar{z})
\Big|_{z=\bar{z}=1/2},
\end{equation}
with $\mathbf{r}$ denoting the representation and $j$ enumerating the different sum-rules.
The components $\vec{V}^{j}_{\mathbf{r},\Delta,\ell}$ can be read off from the columns of the 
sum rules in the different representations. 
For example, in the case of $SU(2)$, in the $(S,\bar{S})_s$ representation, one has from (\ref{eq:SU2sr1})-(\ref{eq:SU2sr3})
\begin{equation}
\vec{V}_{(S,\bar{S}_s),\Delta,\ell} = \left(
\begin{array}{c}
F_{(S,\bar{S})_s,\Delta,\ell} \\
\frac{2}{3}F_{(S,\bar{S})_s,\Delta,\ell} \\ 
\frac{10}{3}H_{(S,\bar{S})_s,\Delta,\ell}
\end{array}\right).
\end{equation}
Instead of minimizing over $a_{mn}$, with
$m$ and $n$ labeling the number of $z$ and $\bar{z}$ derivatives in the linear functional, respectively, we now minimize over $a_{mn}^\mathbf{r}$.
One now writes down a positivity constraint for each representation appearing with each spin.
Note that the number of structures is equal to the number of sum-rules.

In principle, any integer spin and any conformal dimension $\Delta$ satisfying the unitarity bound can appear in the spectrum.
Recall also that real supermultiplets are limited to having $\Delta \geq 2$, rather than the unitarity bound $\Delta \geq 1$,
due to the unitarity constraint on their current superdescendant (cf. 2.3 of \cite{Poland2011}).
Solving the problem numerically requires one to reduce the number of constraints to finite size, which was done, for example, in \cite{Rattazzi:2008pe, Poland:2010wg,Rattazzi:2010gj} by
discretizing the continuous parameter $\Delta$ and setting an upper limit on the spins and scaling dimension for each spin.
This essentially reduces the system to a finite-dimensional linear programming problem, and the minimization can be solved by known algorithms. Notice that such a reduction necessarily omits the constraints for high spins and scaling dimensions,
and the resulting bound may be invalid if it violates these constraints.
In order to somewhat alleviate this concern, one can check that the constraints are not violated at high spins and conformal dimensions using the known asymptotics of the conformal blocks.

We have implemented the computation described in Appendix B of \cite{Poland:2010wg}.
The size of the search space was varied between $k=6$ and $k=8$. We have restricted to the set of spins $\ell = 0,1,...30,100,101$.
Above each unitarity bound $\Delta_{min}$ we consider the following set of dimensions
\begin{equation}
D = \left\{ \Delta_{min} + n \epsilon : n = 0,...,N \right\}
\end{equation}
with $\epsilon = 0.05$ and $N\epsilon = \Delta_{max} - \Delta_{min} = 50$. 
The linear programming was set up using {\tt Mathematica 8.0} and was solved using the Barrier optimizer of {\tt IBM ILOG CPLEX\footnote{http://www-01.ibm.com/software/integration/optimization/cplex-optimizer/}}, which 
is well-suited for large, sparse problems.

\bibliographystyle{JHEP}
\end{document}